\documentclass[prd,aps,twocolumn,amsmath,amssymb,nofootinbib,preprintnumbers]
{revtex4}

\usepackage{graphicx}
\usepackage{dcolumn}
\usepackage{bm}
\usepackage{amsmath}
\usepackage{amsthm}

\usepackage{amsfonts}
\usepackage{euscript,bbm}
\usepackage{ifthen}
\usepackage{psfrag}
\usepackage{slashed}

\usepackage{enumitem}
\usepackage{subfig}

\def\ls{\mathrel{\lower4pt\vbox{\lineskip=0pt\baselineskip=0pt
           \hbox{$<$}\hbox{$\sim$}}}}
\def\gs{\mathrel{\lower4pt\vbox{\lineskip=0pt\baselineskip=0pt
           \hbox{$>$}\hbox{$\sim$}}}}
\def\drawbox#1#2{\hrule height#2pt

\hbox{\vrule width#2pt height#1pt \kern#1pt
              \vrule width#2pt}
              \hrule height#2pt}

\def\Asym#1#2{\vcenter{\vbox{\drawbox{#1}{#2}
              \kern-#2pt       
              \drawbox{#1}{#2}}}}

\newcommand{\be}{\begin{equation}}
\newcommand{\ee}{\end{equation}}
\newcommand{\bea}{\begin{eqnarray}}
\newcommand{\eea}{\end{eqnarray}}
\newcommand{\bi}{\begin{itemize}}
\newcommand{\ei}{\end{itemize}}

\newcommand{\neu}[1]{\ensuremath{\tilde{\chi}_{#1}^0}}

\newcommand{\chpm}[1]{\ensuremath{\tilde{\chi}_{#1}^{\pm}}}

\newcommand{\st}{\ensuremath{\tilde{t}}}

\newcommand{\slep}{\ensuremath{\tilde{\ell}}}

\newcommand{\gsim}{\lower.7ex\hbox{$\;\stackrel{\textstyle>}{\sim}\;$}}
\newcommand{\lsim}{\lower.7ex\hbox{$\;\stackrel{\textstyle<}{\sim}\;$}}

\newcommand{\tanb}{\ensuremath{\tan\beta}}

\newcommand{\ttbar}{t \bar{t}}

\newcommand{\met} {{E\!\!\!\!/_{\rm T}}}

\newcommand{ \pythia } {{\tt PYTHIA}}

\newcommand{ \madgraph } {{\tt MADGRAPH5}}
\newcommand{ \delphes } {{\tt DELPHES}}
\newcommand{ \suspect } {{\tt SuSpect}}

\begin{document}

%
\title{Natural Supersymmetry, Muon $g-2$, and the Last Crevices for the Top Squark}

\author{B. Paul Padley$^{1}$}
\author{Kuver Sinha$^{2}$}
\author{Kechen Wang$^{3}$}

\affiliation{$^{1}$~Physics \& Astronomy Department, Rice University, Houston, TX, 77005, USA \\
$^{2}$~Department of Physics, Syracuse University, Syracuse, NY 13244, USA \\
$^{3}$~Center for Future High Energy Physics, Institute of High Energy Physics, Chinese Academy of Sciences, Beijing, 100049, China
}

\begin{abstract}

We study the interplay of natural supersymmetry and a supersymmetric solution to the discrepancy observed in measurements of the muon magnetic moment. The strongest constraints on the parameter space currently  come from chargino searches in the trilepton channel and slepton searches in the dilepton channel at the LHC, and vast regions are currently allowed, especially at large $\tan{\beta}$. With light top squarks in the spectrum, as required from naturalness arguments, the situation changes dramatically; stop-assisted chargino and neutralino production via $\st \rightarrow b \chpm{1}$ and $\st \rightarrow t \neu{1}$ are already probing the entire parameter space compatible with the muon magnetic moment at $\tan{\beta} \sim \mathcal{O}(10)$, while upcoming stop searches will probe most of the parameter space at larger $\tan{\beta} \sim 40$. Direct Higgsino searches as well as compressed slepton and stop searches are crucial to close out corners of parameter space. We consider one such example: in the presence of light sleptons and charginos as required to obtain appreciable contributions to the muon magnetic moment, compressed stops can dominantly undergo the following decay $\st \, \rightarrow \, b \tilde{\ell} \nu (\tilde{\nu} \ell) \, \rightarrow \, b \ell \nu \neu{1}$, facilitated by off-shell charginos. We find that the enhanced branching to leptons leads to a $5 \sigma$ mass reach (with $3000$ fb$^{-1}$ of data at LHC14) of $m_{\st} \, \sim \, 350$ GeV, with the mass difference between stops and the lightest neutralino being $\sim 80$ GeV. This will further close out a significant part of the parameter space compatible with naturalness and the muon magnetic moment.

\end{abstract}
\maketitle


\section{Introduction}

The anomalous magnetic moment of the muon, $a_\mu = (g-2)_\mu/2$, is one of the most precisely measured quantities in particle physics \cite{Bennett:2006fi}. It will soon be measured to even higher degrees of precision by the New $g-2$ Experiment at Fermilab \cite{Grange:2015eea} and the J-PARC experiment in Japan \cite{jparc_ref}. Since its theoretical value within the Standard Model (SM) can be calculated to within sub-parts-per-million precision, a comparison between theory and experiment provides a powerful probe of new physics. In fact, there has been a long-standing discrepancy between the measured and SM values, denoted by $\Delta a_{\mu}$, given by
\be
\Delta a_{\mu} \, \equiv \, a_\mu{\rm (exp)} \, - \,  a_\mu {\rm (SM)} \, = \, (26.1 \pm 8.0 ) \times 10^{-10} \,\,.
\ee
The above value uses \cite{Hagiwara:2011af} for contributions of the hadronic vacuum polarization, and \cite{Prades:2009tw} for the hadronic light-by-light contribution. The theoretical calculation by \cite{Davier:2010nc} gives the discrepancy to be $\Delta a_{\mu} \, = \, (28.7 \pm 8.0 ) \times 10^{-10}$. Either way, one has a $\sim \, 3\sigma$ deviation from the SM value, providing a tantalizing possibility of new physics.

Supersymmetry provides a particularly appealing framework to understand this anomaly \cite{Martin:2001st}. The size of the discrepancy is comparable to the SM Electroweak contribution $a_\mu {\rm (EW)} \, \sim \, (15.36 \pm 0.1) \times 10^{-10}$ \cite{Czarnecki:2002nt, Czarnecki:1995sz, Czarnecki:1995wq}. Parametrizing new physics contributions to $\Delta a_{\mu}$ by a coupling constant $\alpha_{new}$ and a new physics scale $\Lambda_{new}$ in the loop, this implies that
\bea
\Delta a_\mu \, &\sim &\, \frac{\alpha_{new}}{4 \pi} \frac{m^2_\mu}{\Lambda^2_{new}}\,; \nonumber \\
\alpha_{new} \, \sim \, \alpha_{EW},&& \,\,\,\,\,\,\,\, \,\,\, \Lambda_{new \, \equiv \, SUSY?} \, \sim \, \Lambda_{EW} \,\,.
\eea
Much like the dimensional arguments leading to the so-called dark matter WIMP miracle, the fact that the muon $g-2$ anomaly prefers new physics at the Electroweak scale makes supersymmetry a prime contender in its resolution. The parameter space of the Minimal Supersymmetric Standard Model (MSSM) has been widely studied in this context \cite{Martin:2001st}, \cite{Feng:2001tr} \footnote{Although our focus will be supersymmetry, we point out that the muon magnetic moment has been studied in other contexts recently, for example in \cite{Queiroz:2014zfa}.}.

 The supersymmetric particles that are most relevant for significant contributions to the $\mu \mu \gamma$ vertex (smuons, muon sneutrinos, charged and neutral Higgsinos and Winos, Binos) all have small direct production cross-section at the Large Hadron Collider (LHC). Hence, it is not a surprise that the first run has not yielded any evidence for a supersymmetric resolution of the muon $g-2$ anomaly. The situation, however, is considerably more serious if one considers the main motivation for supersymmetry - stabilizing the gauge hierarchy. One can then advance fine-tuning arguments to make the case that color-assisted production of these particles should already have probed a substantial part of the available parameter space. The quadratic divergence in the Higgs mass (coming dominantly from the top sector of the SM) is cancelled in supersymmetric extensions by loops of scalar partners of the top quark, the top squarks or stops. Stabilizing the Higgs mass with small fine-tuning thus requires light stops. Since the  Electroweak symmetry breaking condition within the MSSM shows that the Higgs vacuum expectation value (vev) cannot be much larger than the Higgsino mass parameter $\mu$, natural Electroweak symmetry breaking also requires light Higgsinos. This is the framework of natural supersymmetry \cite{Feng:2013pwa} \footnote{On the other hand, it is the stop sector that provides the dominant radiative corrections to the Higgs quartic coupling; obtaining the measured value of 125 GeV for the Higgs mass within the MSSM thus requires stops above a TeV, unless the $A$-terms are tuned to maximal mixing. This tension between the fine-tuning associated with obtaining a 125 GeV Higgs, and at the same time obtaining natural Electroweak symmetry breaking, is the little hierarchy problem. Assuming that the little hierarchy problem can be solved through some extension of the MSSM, it is reasonable to expect a natural spectrum. }. 

In this paper, we investigate the interplay between a natural supersymmetric spectrum and the possibility of addressing the muon $g-2$ anomaly within supersymmetry, in the backdrop of constraints coming from the first run of the LHC. We should clarify that we do not consider a natural spectrum in the strict sense that all superpartners that don't play a role in Electroweak symmetry breaking are decoupled; that would exclude light sleptons, for example. Rather, we search for regions of parameter space that would be compatible with the observed discrepancy of the muon magnetic moment and the principal components of a natural spectrum - light Higgsinos and light stops. 

Contributions to the muon magnetic moment in the MSSM are enhanced in two cases: the chargino/sneutrino diagrams and the Bino/smuon diagrams. These contributions are constrained mainly by chargino searches at the LHC in trilepton final states (in the case that the chargino being directly produced is a charged Wino) and by slepton searches in dilepton final states. Broadly, we find that the combination of these searches is only able to rule out a modest part of the parameter space for $\tan{\beta} \, \sim \, \mathcal{O}(10)$, while for larger $\tan{\beta} \sim 40$, the parameter space compatible with $\Delta a_{\mu}$ is essentially unconstrained now, and will depend on the future performance in these two channels at 14 TeV.

Taking into acount color-assisted production of these particles from light stops changes the situation dramatically. Stop searches in the $\st \rightarrow b \chpm{1}$ channel are probing the entire parameter space in case of chargino/sneutrino contributions for $\tan{\beta} \sim \mathcal{O}(10)$, and much of the parameter space for $\tan{\beta} \sim 40$. Similarly, stop searches in the $\st \rightarrow t \neu{1}$ channel are probing much of the parameter space in case of Bino/smuon contributions, for a variety of choices of other parameters. The performance of the LHC in these two stop search channels at 14 TeV will probe much of the remaining parameter space compatible with natural supersymmetry and muon $g-2$. This presents a sobering perspective on the prospects of addressing both the gauge hierarchy and the discrepancy in muon $g-2$ within minimal supersymmetry.

 If one is willing to allow the fine-tuned possibility that the mass difference between the stop and the lightest supersymmetric particle (LSP) is less than the top mass (stop "three-body decay region") or even less than the $W$ mass (stop "four-body decay region"), then the constraints on stops are considerably weaker and much of the parameter space available for the muon magnetic moment opens up, as the bounds revert exclusively to the chargino and slepton searches. Our second set of main results is to probe this important caveat: the presence of light, compressed stops which could have evaded observation. In the presence of light charginos and sleptons, an important decay mode of the stop opens up: $\st \, \rightarrow  \, b \tilde{\chi}^{\pm (*)}_1 \, \rightarrow \,b \tilde{\ell} \nu (\tilde{\nu} \ell) \, \rightarrow \, b \ell \nu \neu{1}$. This decay can be competitive or even dominate over conventional 3-body or 4-body decays of stops through on-shell or off-shell $W$'s. We study this decay in the compressed regions, where $ 175$ GeV $ > (m_{\st} - m_{\neu{1}}) \, \gsim $ 85 GeV, and  $ 85$ GeV $ \gsim (m_{\st} - m_{\neu{1}}) \, \gsim $ 0 GeV. We find that the 14 TeV LHC can discover stops in this decay mode up to a mass of $\sim 350$ GeV at a luminosity of $3000$ fb$^{-1}$, when the mass difference of the stop with the LSP is $\sim 80$ GeV. This will serve the function of further closing out the muon $g-2$ parameter space from the point of view of naturalness.

The rest of the paper is structured as follows. In Section \ref{susymuongminus2}, we describe the features of the parameter space in the MSSM that give rise to appreciable contributions to the muon magnetic moment. In Section \ref{collconstr}, we summarize existing collider constraints on the superpartners that are relevant to our study. In Section \ref{mssmparaspace}, we give our first set of main results, showing the existing collider constraints and top squark probes in regions of parameter space that are compatible with the observed value of $\Delta a_{\mu}$. In Section \ref{discussprospects}, we discuss the results from the parameter space analysis. In Section \ref{stop2slepanalysis}, we present our second set of main results, analysing the collider prospects of compressed stops decaying to the lightest neutralino via sleptons. We end with our Conclusions.

\section{Muon $g-2$ in Supersymmetry} \label{susymuongminus2}

\begin{figure}[!htp]
\centering
\includegraphics[width=4.0in]{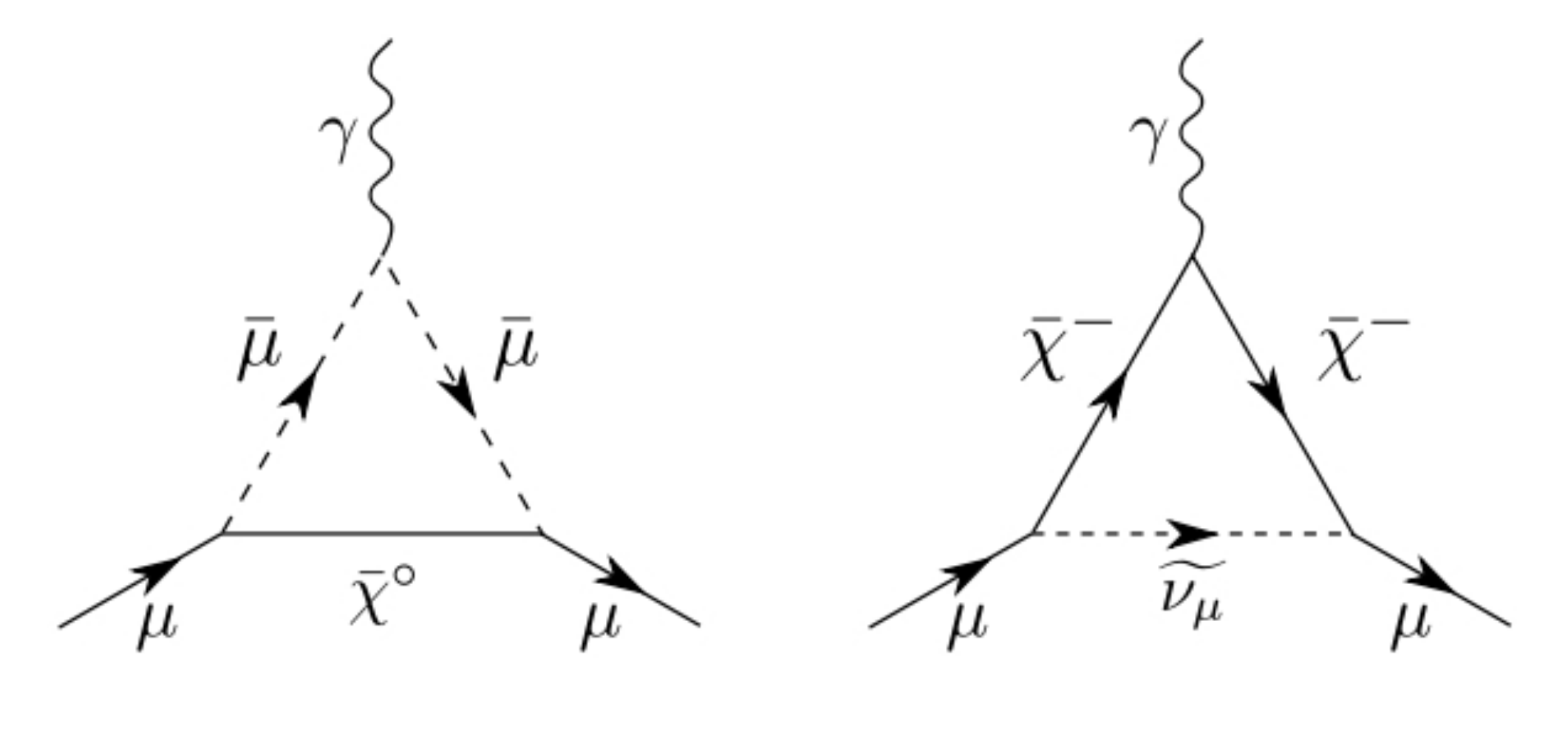}
\caption{Feynman diagrams contributing to muon $g-2$ at one-loop in the MSSM. The left diagram shows the case involving a neutralino and a smuon, while the right diagram shows the case involving a chargino and a muon sneutrino.}
\label{MSSM_muon_gminus2}
\end{figure}

The MSSM  parameter space can easily accommodate the observed discrepancy of the muon magnetic moment. The one-loop contributions are well known and displayed in Fig. \ref{MSSM_muon_gminus2}. The left panel shows the case where a smuon and a neutralino dominate the one-loop contribution. In this case, the charginos can be heavy. On the other hand, the right panel shows the case where the contribution is driven by a light chargino and a muon sneutrino. Since the LSP has to be a neutralino, this means that the spectrum in this case has a light neutralino, in addition to a chargino and the muon sneutrino.

Both classes of contributions can be probed by studying the space spanned by the following parameters:
\be
\mu, M_1, M_2, m_{\tilde{\mu}_L}, m_{\tilde{\mu}_R}, m_{\tilde{\nu}_{\mu}}, \tan{\beta} \,,\label{g2pars} 
\ee
where $\mu$ is the Higgsino mass parameter, $\tan{\beta}$ is the ratio of the Higgs vev's, 
$M_1$ and $M_2$ are the Bino and Wino masses, $m_{\tilde{\mu}_L}$ and $m_{\tilde{\mu}_R}$ are smuon masses, and $m_{\tilde{\nu}_{\mu}}$ is the soft mass of the muon sneutrino.  

The analytic expressions for these diagrams are given below \cite{Kosower:1983yw,Yuan:1984ww, Moroi:1995yh, Martin:2001st, Fargnoli:2013zia}.
\bea
\Delta_{\chpm{1} \tilde{\nu}_{\mu}}&=&\frac{g^2}{(4 \pi)^2}\frac{m_{\mu}^2\tanb}{\mu M_2}\,\mathcal{F}_{[\chpm{1} \tilde{\nu}_{\mu}]}\left(\frac{\mu^2}{m^2_{\tilde{\nu}_{\mu}}},\frac{M^2_2}{m^2_{\tilde{\nu}_{\mu}}}\right)\,,\label{charsneu}\\
\Delta^{(1)}_{\chi\, \tilde{\mu}}&=&-\frac{1}{2}\frac{g^2}{(4 \pi)^2}\frac{m_{\mu}^2\tanb}{\mu M_2}\,\mathcal{F}_{[\chi\, \tilde{\mu}]}\left(\frac{\mu^2}{m^2_{\tilde{\mu}_L}},\frac{M^2_2}{m^2_{\tilde{\mu}_L}}\right)\,,\\
\Delta^{(2)}_{\chi\, \tilde{\mu}}&=&\frac{1}{2}\frac{g'^2}{(4 \pi)^2}\frac{m_{\mu}^2\tanb}{\mu M_1}\,\mathcal{F}_{[\chi\, \tilde{\mu}]}\left(\frac{\mu^2}{m^2_{\tilde{\mu}_L}},\frac{M^2_1}{m^2_{\tilde{\mu}_L}}\right)\label{negl1}\,,\\
\Delta^{(3)}_{\chi\, \tilde{\mu}}&=&-\frac{g'^2}{(4 \pi)^2}\frac{m_{\mu}^2\tanb}{\mu M_1}\,\mathcal{F}_{[\chi\, \tilde{\mu}]}\left(\frac{\mu^2}{m^2_{\tilde{\mu}_R}},\frac{M^2_1}{m^2_{\tilde{\mu}_R}}\right)\label{negl2}\,,\\
\Delta^{(4)}_{\chi\, \tilde{\mu}}&=&\frac{g'^2}{(4 \pi)^2}\frac{m_{\mu}^2 M_1 \mu}{m^2_{\tilde{\mu}_L} m^2_{\tilde{\mu}_R}} \tanb\,\mathcal{F}_{[\chi\, \tilde{\mu}]}\left(\frac{m^2_{\tilde{\mu}_R}}{M^2_1},\frac{m^2_{\tilde{\mu}_L}}{M^2_1}\right)\,,\label{neutsmu}
\eea
where $g$ and $g'$ are the gauge couplings of the $SU(2)$ and $U(1)$ SM groups, respectively, and the $\mathcal{F}_{[\chpm{1} \tilde{\nu}_{\mu}]}$ and $\mathcal{F}_{[\chi\, \tilde{\mu}]}$ are loop functions given by
\bea 
\mathcal{F}_{[\chpm{1} \tilde{\nu}_{\mu}]}(x,y) \, = \, \qquad \qquad \qquad \qquad \qquad \qquad \qquad \qquad \qquad \qquad   \nonumber \\
= xy\left\{\frac{5-3(x+y)+xy}{(x-1)^2(y-1)^2}-  \frac{2}{x-y}\left[\frac{\ln x}{(x-1)^3}-\frac{\ln y}{(y-1)^3}\right]\right\}\,, \nonumber \\
\mathcal{F}_{[\chi\,\tilde{\mu}]}(x,y) \, = \, \qquad \qquad \qquad \qquad \qquad \qquad \qquad \qquad \qquad \qquad   \nonumber \\
= xy\left\{\frac{-3+x+y+xy}{(x-1)^2(y-1)^2}+\frac{2}{x-y}\left[\frac{x\ln x}{(x-1)^3}-\frac{y\ln y}{(y-1)^3}\right]\right\}\,. \nonumber 
\eea
In the above, the reduced forms of \cite{Endo:2013bba} have been used. Also, $\Delta_{\chpm{1} \tilde{\nu}_{\mu}}$ denotes contributions from chargino/sneutrino diagrams, while $\Delta^{(1)}_{\chi\, \tilde{\mu}}, \Delta^{(2)}_{\chi\, \tilde{\mu}}, \Delta^{(3)}_{\chi\, \tilde{\mu}}, $ and $\Delta^{(4)}_{\chi\, \tilde{\mu}}$ denote the contributions from neutralino/smuon diagrams.

Several features of the above equations should be noted:

\begin{itemize}

\item Larger $\tan{\beta}$ values will result in the most conservative collider bounds on the superpartners. This is because large $\tan{\beta}$ enhances the supersymmetric contribution to the magnetic moment. This enables the superpartners participating in the diagrams to have higher masses, where the collider bounds are weaker. In our study, we will generally use $\tan{\beta} \, = \, 40$.

\item The chargino-sneutrino contributions are suppressed for large $\mu$, i.e. heavy Higgsinos. In fact, we will find later that the optimal contributions from these diagrams to $\Delta a_{\mu}$ come from Higgsinos with mass $\lsim \, \mathcal{O}(500)$ GeV, which is precisely the region preferred by small Electroweak fine-tuning arguments. For heavier Higgsinos, either the charged Wino has to be light, or the neutralino-smuon contributions of Eq.~\ref{neutsmu} have to dominate.

\item For the neutralino/smuon contributions, either the left-handed or the right-handed smuon has to be light, while the other can be heavy. This has some implications for collider bounds, which are typically more severe for left-handed sleptons. 

\item In the limit of heavy charginos, the neutralino-smuon contribution of Eq.~\ref{neutsmu} dominates. Again, either the left or right-handed smuon has to be light; depending on how heavy the charged Wino and Higgsino are, they may both have to be light.

\end{itemize}

In the next Section, we go on to discuss the relevant collider bounds coming from the LHC, before applying the bounds on the MSSM parameter space compatible with $\Delta a_{\mu}$ in Section \ref{mssmparaspace}.

\section{Collider Constraints} \label{collconstr}

In this section, we discuss the limits on supersymmetric spectra obtained from LEP and LHC searches. This will enable us to conveniently refer back to the relevant search when we discuss the parameter space in the next section. We will concentrate on the particles relevant for us, namely $\neu{1}$, $\tilde{\ell}$, $\st$, and $\chpm{1}$.

\subsubsection*{LEP bounds:}

LEP results ~\cite{LEPref} constrain slepton and chargino masses: 
\begin{itemize}

\item sleptons: $m_{\tilde{\ell}_L}, m_{\tilde{\ell}_R} \, > \, 100.0 $ GeV where $\ell = e,\mu$; ,

\item lightest chargino : $m_{\chpm{1}} \, > \, $ 103.5 GeV.

\end{itemize}

\subsubsection*{LHC bounds:}

The search for gluinos ($\tilde{g}$) and first two generation squarks ($\tilde{q}$) at the Large Hadron Collider (LHC) has so far yielded null results. The exclusion limits on squark  and gluino  masses, when they are comparable, are approximately $1.5$ TeV at $95\%$ CL with $20$ fb$^{-1}$ of integrated luminosity \cite{:2012rz, Aad:2012hm, :2012mfa, LHCsquarkgluino20ifb}. We will assume, in the remainder of the paper, that these states are decoupled from the low energy physics. Therefore, in studying the constraints on the superpartners relevant for our study, we will not consider the limits arising from gluino/squark production followed by cascade decay to $\neu{1}, \, \tilde{\ell},$ or $\chpm{1}$. As discussed in the Introduction, cascade decays of the top squark will be important for us.


\begin{enumerate}[label= (\Roman*), leftmargin = 0.4 cm]

\item

{\it {\bf Chargino and neutralino masses -}} Limits on the lightest charginos coming from direct chargino production or production from top squarks are as follows.
\begin{enumerate}
\item We first consider the case where the Bino is the LSP, and there is no light slepton in the spectrum, i.e., $m_{\tilde{\ell}} \, > \, m_{\chpm{1}}, m_{\neu{2}}$, where the neutral and charged Winos constitute $\neu{2}$ and $\chpm{1}$, respectively. For direct charged Wino production, the signatures are $WH + \met$ or $WZ + \met$ final states, with model dependent branching fractions. Taking into account both ATLAS and CMS results, one can rule out charged Wino masses below $350$ GeV for $\neu{1} \, \lsim $ 140 GeV \cite{Aad:2014nua}. 
\item We next consider the case where the Bino is the LSP, and there is a light slepton in the spectrum, i.e., $m_{\tilde{\ell}} \, < \, m_{\chpm{1}}, m_{\neu{2}}$, where the neutral and charged Winos constitute $\neu{2}$ and $\chpm{1}$, respectively. For direct charged Wino production with a lighter slepton, the exclusion bounds are much stronger, coming from trilepton searches \cite{Aad:2014nua, Khachatryan:2014qwa}. For $\neu{1} \, \lsim $ 300 GeV, $\chpm{1}$ is excluded below 720 GeV \cite{Aad:2014nua, Khachatryan:2014qwa} assuming sleptons intermediate to $\chpm{1}$ and $\neu{1}$. Compressed regions of paramemter space with $m_{\chpm{1}} - m_{\neu{1}} \, \lsim \, 30$ GeV are still allowed due to insufficient $\met$. We will use Figures $7(a)$, $7(b)$, $7(c)$, and $7(d)$ of \cite{Aad:2014nua} for our work.

\item For cases where the neutral and charged Winos themselves comprise the LSP and next-to-lightest supersymmetric particles, the bounds are much weaker and it is useful to think in terms of high luminosity studies at 14 TeV. Taking a background systematic uncertainty of $1\% (5\%)$, with $3000$ fb$^{-1}$ of data, the 14 TeV LHC is sensitive to Winos of 240 GeV (125 GeV) and Higgsinos of 125 GeV (55 GeV) \cite{Delannoy:2013ata, Berlin:2015aba, Cirelli:2014dsa} in a pure cut and count analysis. Shape analysis can improve the mass reach significantly \cite{Delannoy:2013ata}.

\item We now consider bounds coming from direct stop production, in spectra with $m_{\st} \, > \, m_{\chpm{1}}, m_{\neu{1}}$. For charginos produced on-shell from top squarks, there are very stringent limits depending on the relative mass separations of $\neu{1}$, $\chpm{1},$ and $\st$. For large mass separations between the lightest chargino and the LSP, $\chpm{1} \, = \, 2 \neu{1}$, direct production of stops followed by the decay $\st \rightarrow b \chpm{1}$ rules out stops up to $\sim \, 500$ GeV with $\chpm{1} \, \lsim 400$ GeV \cite{Aad:2014kra}, \cite{Aad:2014qaa}.  

\item For $m_{\st} \, > \, m_{\chpm{1}}, m_{\neu{1}}$ and for compressed chargino-neutralino system with $m_{\chpm{1}} - m_{\neu{1}} \, \sim \, 5$ GeV, direct production of stops followed by the decay $\st \rightarrow b \chpm{1}$ rules out stop masses up to $m_{\st} \, \sim \, 600$ GeV for $m_{\neu{1}} \, \sim \, 200$ GeV, unless chargino-stop is also compressed, with a mass separation of $\sim 40$ GeV \cite{Aad:2014kra}. 
For compressed chargino-neutralino system but with $m_{\neu{1}} \, \sim \, 200 - 250$ GeV, top squarks are ruled out up to $600$ GeV except for a window $250-350$ GeV.
\end{enumerate}
%


\item

{\it {\bf Slepton masses -}} The limits on sleptons are particularly strong when the initial particle in the decay chain is a squark/gluino or a stop or chargino, as discussed previously. However, for direct production of sleptons, the limits are much weaker due to small production cross section. 

\begin{enumerate}

\item

 The direct production of sleptons has been probed both by dilepton searches at ATLAS \cite{ATLASSlep} and CMS \cite{CMSSlep1}. The decay chain is $pp \rightarrow \slep \slep^* \rightarrow l^+ l^- \neu{1} \neu{1}$, with $Br(\slep \rightarrow l^- \neu{1}) = 1$.; the final states containing opposite-sign same-flavor non-resonant dileptons and missing transverse energy ($\met$).  The mass separation $ m_{\tilde{\ell}}-m_{\neu{1}}$ is an important factor in the exclusion plots. The mass reach with $m_{\neu{1}} = 0$ GeV is $m_{\tilde{\ell}} \sim 280$ GeV and 330 GeV for CMS and ATLAS respectively. We will use Figures $8(a), \, 8(b), \,$ and $8(c)$ of \cite{ATLASSlep} for our analysis.
 

\item

Compressed spectra with smaller mass separation between $\tilde{\ell}$ and $\neu{1}$ may have eluded these probes. For mass difference of $\tilde{\ell}$ and $\neu{1}$ of 5-15 GeV, a recent study with VBF tagged jets and $\met$ has found that the 14 TeV LHC may be sensitive to sleptons up to mass $m_{\tilde{\ell}}$=115-135 GeV, with 3000 fb$^{-1}$ of data \cite{Dutta:2014jda}. Compressed slepton searches have been studied by several other groups as well \cite{Han:2014aea}.

\end{enumerate}


\item

{\it {\bf Stop masses -}} The bounds on the mass of the lightest top squark ($\st$) are less stringent than those on gluino or first two generation squark masses, due to smaller production cross section. Exclusion limits in the $m_{\st}$-$m_{\neu{1}}$ plane have been obtained for a variety of decay modes:
\begin{enumerate}[leftmargin=*]

\item   The simplest scenario for $\st$ studies is to consider the direct QCD production of $\st$ pairs with $100 \%$ branching $\st \rightarrow t \neu{1}$. Exclusion limits in the $m_{\st}$-$m_{\neu{1}}$ plane have been obtained in this decay mode (monolepton search \cite{Aad:2014kra}, dilepton search \cite{Aad:2014qaa}). The limits depend on the mass difference $m_{\st} - (m_{\neu{1}} + m_t)$. For $m_{\neu{1}} \, \lsim \, 140$ GeV, stop masses up to $m_{\st} \, \sim \, 650$ GeV have been ruled out, except for highly compressed scenarios $m_{\st} - (m_{\neu{1}} + m_t) \, \sim \, 20$ GeV. For $m_{\neu{1}} \, \lsim \, 140 - 200$ GeV, the same limits of stop masses apply, except $m_{\st}$ in the interval $320$ - $420$ GeV also opens up. 

\item Three body stop decays into $bW \neu{1}$ have also been bounded between the limits $m_{\st} \, > \, m_b + m_W + m_{\neu{1}}$ and $m_{\st} \, < \, m_t + m_{\neu{1}}$. These constraints are weaker due to smaller $\met$ and softer $b$-jets. The limits extend up to $m_{\st} \, \sim \, 300$ GeV and $m_{\neu{1}} \, \sim \, 120$ GeV. While most of the parameter space in this region has been ruled out, regions close to the boundary $m_{\st} \, = \, m_b + m_W + m_{\neu{1}}$ are still allowed.

\item Four body decays of stops satisfying $m_{\st} \, < \, m_b + m_W + m_{\neu{1}}$ have been studied, and the current constraints reach $m_{\st} \, \sim \, 240$ GeV and $m_{\neu{1}} \, \sim \, 160$ GeV.

\end{enumerate}

\end{enumerate}

\subsection{Reinterpreting Slepton and Chargino Bounds for Higgsino LSP} \label{reinterprethsino}

In the LHC constraints for charginos (trilepton channel) and sleptons (dilepton channel) described above, the LSP has been assumed to be a pure Bino. In describing scenarios compatible with natural supersymmetry, we will often be interested in scenarios where the LSP is a Higgsino \cite{hsinoLSP}. The bounds should therefore be reinterpreted for that case.

The main difference (for left-handed sleptons) will come from the fact that now the sneutrino can have non-zero branching to charged leptons, while the slepton can have non-zero branching to neutrinos. The branchings depend on the Bino or Wino components of $\neu{1}, \neu{2},$ and $\chpm{1}$, and are given by \cite{Eckel:2014dza}
\begin{eqnarray}
\Gamma(\tilde\ell \to\ell \chi_{1,2}^0)&=& C_1 \  
( m_Z\frac{  s_W^2}{M_1 \mp \mu}-m_Z\frac{c_W^2}{M_2 \mp \mu})^2,  \label{eq:slL_chi10}\\
\Gamma(\tilde\ell \to\nu_\ell  \chi_{1}^\pm)&=& C_2 \ 8 c_W^4  (c_\beta+ s_\beta \frac{\mu}{M_2})^2 
( \frac{m_Z}{M_2} )^2,  
\label{eq:slL_chi1pm}\\
\Gamma(\tilde\nu_\ell \to\nu_\ell \chi_{1,2}^0)&=& C_1 \ 
( m_Z\frac{  s_W^2}{M_1 \mp \mu}+m_Z\frac{c_W^2}{M_2 \mp \mu})^2,  \label{eq:snuL_chi10}\\
\Gamma(\tilde\nu_\ell \to \ell  \chi_{1}^\pm)&=& C_2 \ 8 c_W^4  (s_\beta+ c_\beta \frac{\mu}{M_2})^2 
( \frac{m_Z}{M_2} )^2 , 
\end{eqnarray}
where  
\begin{equation}
C_1=(\sin{\beta}\pm \cos{\beta})^2 \frac{1}{16 \pi}\frac{e^2}{4 s_W^2 c_W^2}   \frac{(m_{\cal P}^2 - m_{\cal D}^2)^2}{m_{\cal P}^3}
\end{equation}
and
\begin{equation}
C_2=\frac{1}{16 \pi}\frac{e^2}{4 s_W^2 c_W^2}   \frac{(m_{\cal P}^2 - m_{\cal D}^2)^2}{m_{\cal P}^3} \,\,.
\end{equation}

In the above, $m_{\cal P}$ and $m_{\cal D}$ being the  the parent slepton mass and daughter neutralino/chargino mass, respectively. In the Bino LSP case for which the LHC bounds are given, only the slepton production is considered, while the sneutrino branches completely into missing energy. Now, both slepton and sneutrino productions have to be considered, as well as their respective branchings to charged leptons. We will mainly consider $M_1 \sim M_2$, and $\frac{\mu}{M_2} \sim 0.1-0.5$. In that case, the combined production cross section times branching to charged leptons $\sigma \times Br$ reduces approximately to the Bino LSP case. 

We will thus use the same exclusion bounds, for Constraints $(Ib)$ and $(IIa)$, as those given by the collaborations. We mention, however, that for special choices of $M_1, M_2,$ and their relative signs, the slepton mass reach may be reduced to $\sim 230$ GeV or enhanced to $\sim 470$ GeV \cite{Eckel:2014dza}.

The story for the trilepton searches for charginos is similar. The main point is that the trilepton searches will target the production of  the second lightest chargino $\tilde{\chi}_2^\pm$,  which is the charged Wino. This $\tilde{\chi}_2^\pm$ will then decay to a slepton or a sneutrino, which will then decay down to three possible states (the neutral Higgsino, and the two charged Higgsinos), all of which behave effectively as the LSP since the mass difference between the charged and neutral Higgsinos is small. The relevant decay chains are $\chpm{2} \rightarrow \tilde \ell ^{\pm} \nu (\tilde\nu \ell ^{\pm}) \rightarrow \ell ^{\pm} \nu \neu{1,2} , \ell ^{\pm} \ell ^{\mp} \chpm{1}, \nu \chpm{1}$, where in the last step we have shown all the possible final states.

What is relevant for collider bounds is that the effective branching of the charged Wino to the LSP will change relative  to the canonical case where the LSP is a Bino and the Higgsinos are heavy, which is the case for which the Collaborations have given their results. We have computed the new branchings in the region of parameter space relevant for us following Eq. 9-12, and found that the net branching is similar to the canonical case. We have thus used the bounds from the Collaborations, which are actually a little conservative for our case.

\section{$\Delta a_{\mu}$ and a Natural Spectrum} \label{mssmparaspace}

In this section, we describe the parameter space regions which can account for the observed value of $\Delta a_{\mu}$, in the presence and absence of a top squark in the low energy spectrum.


\begin{figure*}[!t]
\centering
\mbox{\includegraphics[width=\columnwidth]{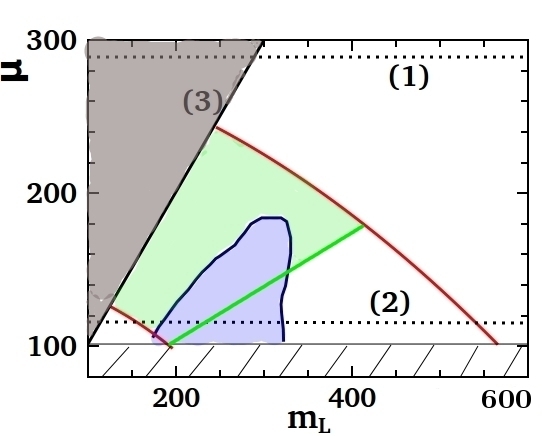}\quad\includegraphics[width=\columnwidth]{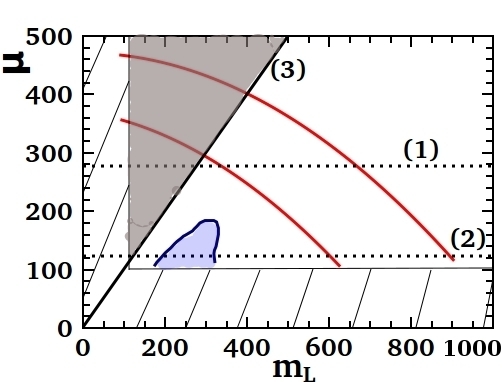}}
\caption{The interplay of the muon $g-2$ anomaly and a natural supersymmetric spectrum, plotted on the $(\mu, m_{\tilde{\ell}_L})$ plane. $M_2 \, = \, 2 \mu$, while $\tilde{\ell}_R$ is kept heavy and decoupled. On the left panel, $\tan{\beta} \, = \, 10$, while on the right panel, $\tan{\beta} \, = \, 40$. The area between the red solid curves is consistent within $2 \sigma$ of the observed value of $\Delta a_\mu$. The blue contour shows the ATLAS dilepton bounds coming from $\tilde{\ell}_L \, \rightarrow \, \ell \neu{1}$, which is Constraint $(IIa)$. The green region shows the part of $\Delta a_{\mu}$ - compatible space that is constrained by the ATLAS trilepton searches coming from  $\chpm{2} \rightarrow \tilde \ell ^{\pm} \nu (\tilde\nu \ell ^{\pm}) \rightarrow \ell ^{\pm} \nu \neu{1,2} , \ell ^{\pm} \ell ^{\mp} \chpm{1}, \nu \chpm{1}$, which is Constraint $(Ib)$ adapted to the case of Higgsino LSP. The hatched region is ruled out by LEP constraints on charginos and sleptons. To the right of the black solid line marked $(3)$, the LSP is a neutralino.  The parameter space below the dotted lines is being probed by current stop limits. The black dotted line marked $(1)$ marks the chargino/neutralino mass constraint for the case of stop-assisted production, which is Constraint $(Ie)$. The black dotted line marked $(2)$ shows the future mass reach for direct chargino/neutralino production in the absence of light stops, with $3000$ fb$^{-1}$ of data at LHC14, which is Constraint $(IIc)$. All masses are in GeV.}
\label{purehsino1}
\end{figure*}

In Fig. \ref{purehsino1}, we show muon $g-2$ constraints for a spectrum containing light Higgsinos, sleptons, and stops. On the left panel, we show the case of $\tan{\beta} \, = \, 10$, with $M_2 \, = \, 2 \mu$. $\tilde{\ell}_R$ is kept heavy and decoupled. The vertical axis plots the Higgsino mass parameter $\mu$, while the horizontal axis plots the mass of the left-handed sleptons $m_{\tilde{\ell}_L}$. We note that $\mu$ sets the scale for both the lightest neutralino LSP, as well as the lightest chargino, which have a mass separation of $\mathcal{O}(5)$ GeV. The hatched region is ruled out from LEP bounds. The region between the red solid lines shows the region of parameter space that is within $2 \sigma$ of the observed value of $\Delta a_{\mu}$ . The blue solid contour shows the limits on the $(m_{\tilde{\ell}_L}, m_{\neu{1}})$ plane coming from the ATLAS dilepton study $\tilde{\ell}_L \, \rightarrow \, \ell \neu{1}$, which is Constraint $(IIa)$. The green region shows the part of $\Delta a_{mu}$ - compatible space that is constrained by the ATLAS trilepton searches coming from charged Wino decay, which is Constraint $(Ib)$. Following our discussion in Sec. \ref{reinterprethsino}, we use the trilepton bounds on charged Winos from ATLAS, even in the case where the Higgsinos and not the Binos constitute the LSP. We note that this is a conservative assumption for the region in parameter space shown in Fig. \ref{purehsino1}.

The left panel of Fig. \ref{purehsino1} has two dotted lines below which the parameter space is being probed by current stop searches: $(1)$ This line shows the chargino/neutralino mass constraint in the case of stop-assisted production, which is Constraint $(Ie)$; $(2)$ This line shows the future mass reach for direct chargino/neutralino production in the absence of light stops, which is Constraint $(Ic)$. In addition, the solid black line marked $(3)$ demarcates the regon where the lightest neutralino is the LSP, as required from dark matter considerations.

It is clear that much of the parameter space required for obtaining the correct value of $\Delta a_\mu$ is already ruled out by the ATLAS dilepton and trilepton studies, while the entire parameter space is in fact being probed in the case of stop-assisted production (except the case of compressed stops).

The right panel of Fig. \ref{purehsino1} shows the same parameter space, with $\tan{\beta} \, = \, 40$ and $M_2 \, = \, 2 \mu$. $\tilde{\ell}_R$ is decoupled as before. In contrast to the left panel, we now see that the enhancement of $\Delta a_\mu$ due to larger $\tan{\beta}$ results in an allowed region between the red solid lines where the LSP and slepton masses are higher. The ATLAS dilepton constraints in blue are now irrelevant, as are the trilepton charged Wino constraints that have not been displayed. There is a small allowed region between $m_{\neu{1}} = 280$ GeV and $m_{\neu{1}} = 400$ GeV which lies above the current bounds coming from a light stop. This region will be explored at Run II of the LHC.

\begin{figure*}[!t]
\centering
\mbox{\includegraphics[width=\columnwidth]{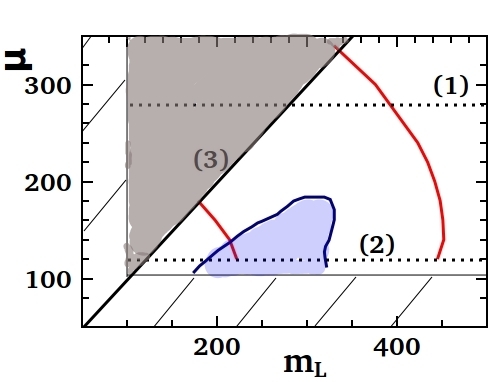}\quad\includegraphics[width=\columnwidth]{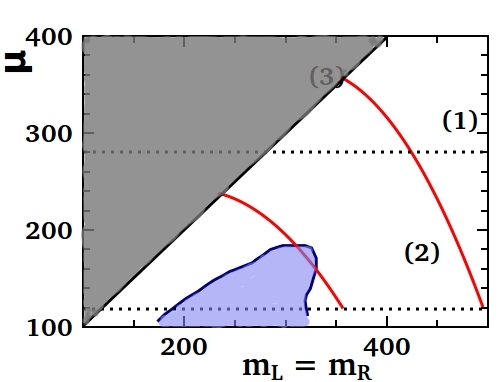}}
\caption{The interplay of the muon $g-2$ anomaly and a natural supersymmetric spectrum, plotted on the $(\mu, m_{\tilde{\ell}_L})$ plane. The fixed parameters are $M_2 \, = \, 1$ TeV, $\tan{\beta} \, = \, 40$. On the left panel, $m_{\tilde{\ell}_R} \, = \, 3$ TeV is heavy and decoupled. On the right panel, $m_{\tilde{\ell}_R} \, = \, m_{\tilde{\ell}_L}$. The legend is the same as Fig. \ref{purehsino1}. The hatched region is ruled out by LEP constraints on charginos and sleptons. All masses are in GeV.}
\label{purehsino2}
\end{figure*}

Fig.\ref{purehsino2} shows the same parameter space as Fig. \ref{purehsino1}, except now $M_2$ is also decoupled from the low-energy spectrum. The left panel shows the case of $\tan{\beta} \, = \, 40$, $M_2 \, = \, 1$ TeV, and $m_{\tilde{\ell}_R} \, = \, 3$ TeV being heavy and decoupled. The right panel shows the case of  $\tan{\beta} \, = \, 40$, $M_2 \, = \, 1$ TeV, and $m_{\tilde{\ell}_R} \, = \, m_{\tilde{\ell}_L}$. The regions between the solid red lines are the allowed space for $\Delta a_\mu$. In the left panel, the ATLAS dilepton bounds rule out a significant portion of the parameter space. The trilepton chargino searches are irrelevant since the charged Wino is too heavy. However, current bounds on stop-assisted production are probing the entire space. The presence of a light right-handed slepton in the right panel allows for larger values of $m_{\neu{1}}$ and $m_{\tilde{\ell}_L}$. The dilepton constraints are mostly irrelevant here, although current bounds on stop-assisted production still probes most of the parameter space. The remaining space above $m_{\chpm{1}} = 280$ GeV will be probed in Run II of the LHC.

\begin{figure*}[!t]
\centering
\mbox{\includegraphics[width=\columnwidth]{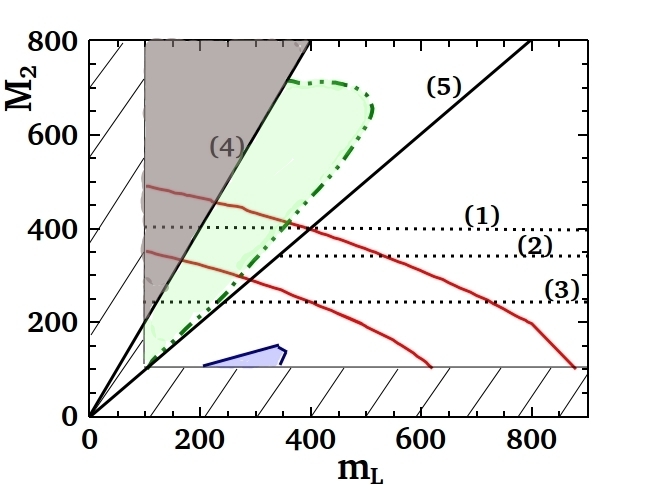}\quad\includegraphics[width=\columnwidth]{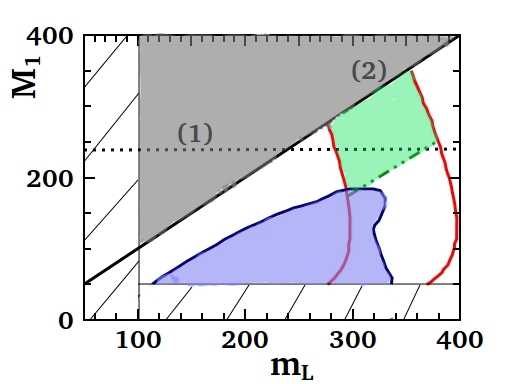}}
\caption{The interplay of the muon $g-2$ anomaly and a natural supersymmetric spectrum, plotted on the $(M_2, m_{\tilde{\ell}_L})$ plane (left panel) and $(M_1, m_{\tilde{\ell}_L})$ plane (right panel), with $\mu$ kept large. {\bf Left panel:} $\tanb = 40$, $\mu = 2 M_2$, and $M_1 \, = \, \frac{1}{2} M_2$. The dot-dashed green contour is the entire region ruled out by the ATLAS trilepton searches with $\chpm{1} \rightarrow \tilde{\ell}^{\pm} \nu (\tilde{\nu} \ell^{\pm}) \, \rightarrow \, \ell^{\pm} \nu \neu{1}$, which is Constraint $(Ib)$, while the solid blue contour is ruled out by ATLAS dilepton searches, which is  Constraint $(IIa)$. To the right of the solid black line marked $(4)$, the LSP is a Bino. To the right of the solid black line marked $(5)$, the slepton is heavier than the charged Wino. The area between the red curves gives the region compatible within $2 \sigma$ of the observed value of $\Delta a_\mu$. The regions below the black dotted lines are being probed by various searches (see text). {\bf Right panel:} $M_2 = 2 M_1$, $\mu = 2$ TeV, and $\tan{\beta} = 40$. The right-handed sleptons are kept fixed at $m_{\tilde{\ell}_R} \, = \, 1.5 \times m_{\tilde{\ell}_L}$. The blue and red contours denote the same regions as the left panel. To the right of the solid black line marked $(2)$, the LSP is a Bino. The dotted black line marked $(1)$ denotes the limits on $m_{\neu{1}}$ coming from $\st \rightarrow t \neu{1}$, which is Constraint $(IIIa)$.  The hatched region is ruled out by LEP constraints on charginos and sleptons. All masses are in GeV.}
\label{Binoslepton1}
\end{figure*}

We next give results for the case where $\mu$ is large, but there is a light stop in the spectrum.  A representative example of the constraints is shown in the left panel of Fig. \ref{Binoslepton1}. We have taken $\tanb = 40$, $\mu = 2 M_2$, and $M_1 \, = \, \frac{1}{2} M_2$. Lower $\tanb$ will make the collider bounds more stringent, since the particles will be pushed to lower masses to account for $\Delta a_\mu$. The LEP excluded regions are hatched. The corridor between the solid red lines indicates the region that is compatible within $2 \sigma$ of the measured value of $\Delta a_\mu$. The green dot-dashed line indicates the full region ruled out from trilepton searches with $\chpm{1} \rightarrow \tilde{\ell}^{\pm} \nu (\tilde{\nu} \ell^{\pm}) \, \rightarrow \, \ell^{\pm} \nu \neu{1}$, which is Constraint $(Ib)$. The blue solid line shows the limits from the ATLAS dilepton study, which is Constraint $(IIa)$ as in the previous figures. For $m_{\slep} \, > \, m_{\chpm{1}}$, the limits are weaker and model-dependent. To the right of the solid black line marked $(4)$ is the region where the LSP is a neutralino, while to the right of the solid black line marked $(5)$ is the region where $m_{\slep} \, > \, m_{\chpm{1}}$. From top to bottom, the dotted black lines give chargino mass bounds for the various scenarios described in the text: $(1)$ top squark $m_{\st} \, > \, m_{\chpm{1}} + m_b$, with decay chain $\st \rightarrow b \chpm{1}$, ruling out $m_{\chpm{1}} \, < \, 400$ GeV, as given by Constraint $(Id)$; $(2)$ decoupled top squark and heavy slepton, with $\chpm{1} \rightarrow Wh \neu{1}$ or $\chpm{1} \rightarrow WZ \neu{1}$, ruling out  $m_{\chpm{1}} \, < \, 350$ GeV for $m_{\neu{1}} \, \lsim $ 140 GeV, which is  Constraint $(Ia)$; $(3)$ top squarks and $M_1$ decoupled, with the neutral Wino being the LSP, which is Constraint $(Ic)$.

On the right panel of Fig. \ref{Binoslepton1}, we show the constraints on the $(M_1, m_{\tilde{\ell}_L})$ plane for the case of $M_2 = 2 M_1$ and $\mu = 2$ TeV, with $\tan{\beta} = 40$. The right-handed sleptons are kept fixed at $m_{\tilde{\ell}_R} \, = \, 1.5 \times m_{\tilde{\ell}_L}$. The region between the red solid lines is allowed by $\Delta a_\mu$. The blue solid contour shows the ATLAS dilepton bound as before, while the green region is ruled out by ATLAS chargino searches. The black dotted line marked $(1)$ shows the limits on $m_{\neu{1}}$ coming from stop decay $\st \rightarrow t \neu{1}$, which is Constraint $(IIIa)$, ruling out the parameter space that is below it. To the right of the solid line marked $(2)$ is the region where the LSP is a neutralino.

\begin{figure*}[!t]
\centering
\mbox{\includegraphics[width=\columnwidth]{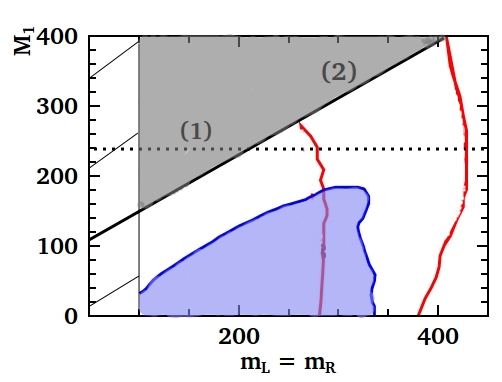}\quad\includegraphics[width=\columnwidth]{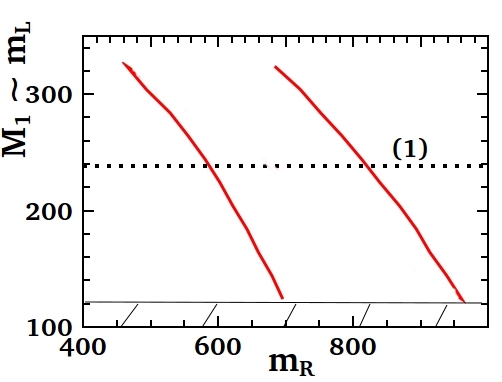}}
\caption{The interplay of the muon $g-2$ anomaly and a natural supersymmetric spectrum, plotted on the $(M_1, m_{\tilde{\ell}_L})$ plane (left panel) and $(M_1, m_{\tilde{\ell}_R})$ plane (right panel), with $\mu$ kept large. On the left panel, $\tan{\beta} = 10$, with $\mu = M_2 = 5$ TeV and $m_{\tilde{\ell}_L} \, = \, m_{\tilde{\ell}_R}$. On the right panel, $\tan{\beta} = 40$, $\mu = 3.3$ TeV, $M_2 = 750$ GeV, and $M_1 \, \sim \, m_{\tilde{\ell}_L}$. The red and blue contours are the same as in previous figures. To the right of the line marked $(2)$ is the region where the LSP is a Bino. Below the dotted black line marked $(1)$, the space is being probed by stop-assisted $\neu{1}$ production, with  limits on $m_{\neu{1}}$ coming from stop decay $\st \rightarrow t \neu{1}$, which is Constraint $(IIIa)$. The hatched region is ruled out by LEP constraints on sleptons. All masses are in GeV.}
\label{Binoslepton2}
\end{figure*}

In Fig. \ref{Binoslepton2}, we continue to show the parameter space in cases where $\mu, M_2$ are heavy. On the left panel, we plot the case of $\tan{\beta} = 10$, with $\mu = M_2 = 5$ TeV. The blue solid contour shows the ATLAS dilepton bound, which is Constraint $(IIa)$. The black dotted line marked $(1)$ shows the limits on $m_{\neu{1}}$ coming from stop decay $\st \rightarrow t \neu{1}$ which is Constraint $(IIIa)$, ruling out the parameter space that is below it. To the right of the solid line marked $(2)$ is the region where the LSP is a neutralino. The right panel shows the same plane, with $\tan{\beta} = 40$, $\mu = 3.3$ TeV, and $M_2 = 750$ GeV. The value of $M_2$ was chosen to correspond to just above the limit where the ATLAS trilepton search, which is  Constraint $(Ib)$, begins to be important.

\section{Prospects: Current and Future} \label{discussprospects}

In the previous sections, we have studied the parameter space for the muon $g-2$ anomaly in the MSSM. The following conclusions can be drawn:

\begin{itemize}

\item If the low energy spectrum consists only of Higgsinos and sleptons, then the current LHC slepton searches rule out a part of the parameter space for $\tan{\beta} \sim \mathcal{O}(10)$, while for larger $\tan{\beta} \sim \mathcal{O}(40)$, the parameter space is completely unconstrained. 

\item If the low energy spectrum consists of light stops in addition to Higgsinos and sleptons, then large portions of parameter space consistent with $\Delta a_{\mu}$ can be probed by stop searches. The entire parameter space consistent with $\Delta a_{\mu}$ for $\tan{\beta} = \mathcal{O}(10)$ has been probed; significant parts have been probed for larger $\tan{\beta} \sim 40$, as we saw for various choices of $M_2$ and $m_{\tilde{\ell}_R}$. The high luminosity LHC will essentially probe this region.

\item In the case of large $\mu$ and light Wino (Fig. \ref{Binoslepton1}) but with no light stops, large parts of parameter space corresponding to regions where $m_{\tilde{\ell}} \, > \, m_{\chpm{1}}$ are allowed, primarily due to the weakness of LHC constraints in that region. Taking into account stop decays, however, most of the parameter space has been probed.

\item For the case of dominant Bino/smuon contributions to $\Delta a_{\mu}$ (Fig. \ref{Binoslepton2}), stops probe a significant part of the parameter space.

\item We note that for the case where the Bino/smuon contribution is dominant and both charginos are heavy, there is no constraint at all from the LHC currently in the large $\tan{\beta} \sim 40$ regime, as is clear from the right panel of Fig. 5. The only constraint comes from a light stop, since the area below the line marked (1) is being probed currently.

\end{itemize}

Current limits on the Electroweak sector coming from the CMS and ATLAS collaborations can only constrain a small part of the avaiable space for $\Delta a_{\mu}$. Taking into account top squark decays, however, the situation changes dramatically. It is clear that a natural supersymmetric spectrum can tell us a lot about the available parameter space for muon $g-2$, and future probes of the top squark are important in this regard.

\subsection{Compressed Stops}

The limits on charginos and  neutralinos coming from stop decay break down if the stop is compressed, i.e., if the total mass of the decay products of the stop is approximately equal to the stop mass. These are the scenarios where there is little $\met$ in the final state. While top squark searches at the next run of the LHC will sweep out larger and larger parts of the parameter space for $\Delta a_{\mu}$, it is important to note that compressed stops provide scenarios that could have evaded detection, and are compatible with natural supersymmetry and muon $g-2$.

The remainder of the paper is devoted to studying such a scenario. The possible compression scenarios for stops are: 

\begin{enumerate}

\item $\st \rightarrow b \chpm{1}$, with $m_{\st} \, \sim \, m_b + m_{\chpm{1}}$, which is important for the natural supersymmetric spectrum in Fig. \ref{purehsino1} and Fig. \ref{purehsino2}, and described in Constraint $(Id)$ and $(Ie)$. This constraint is especially relevant when the chargino/sneutrino contribution to muon $g-2$ is dominant. This decay mode has been studied by several authors \cite{Dutta:2013sta}.

\item  $\st \rightarrow t \neu{1}$, with $m_{\st} \, \sim \, m_t + m_{\neu{1}}$, which is important for Fig. \ref{Binoslepton1}, right panel, and Fig. \ref{Binoslepton2}, where the Bino/smuon contribution to the magnetic moment is dominant. This compressed region has been studied by several authors \cite{Alves:2012ft}, \cite{Dutta:2013gga}.

\item The conventional 3-body and 4-body stop decay regions: $ 175$ GeV $ > (m_{\st} - m_{\neu{1}}) \, \gsim $ 85 GeV, and  $ 85$ GeV $ \gsim (m_{\st} - m_{\neu{1}}) \, \gsim $ 0 GeV, respectively. In these regions, the $b$-jet $p_T$ is typically too small to allow efficient $b$-tagging, in addition to small $\met$. The 3-body region has been explored by \cite{Dutta:2013gga}.

\end{enumerate}

While all of these scenarios are important, we will especially be interested in scenarios that are affected by the presence of light charginos and sleptons, which are required for a large contribution to the muon magnetic moment. In particular, the decay of stops to off-shell charginos 
\be \label{stopslepdecay}
\st \, \rightarrow  \, b \tilde{\chi}^{\pm (*)}_1 \, \rightarrow \,  b \tilde{\ell} \nu (\tilde{\nu} \ell) \, \rightarrow \, b \ell \nu \neu{1}
\ee
can be competitive with decays in the conventional 3-body and 4-body regions, which occur through on-shell and off-shell $W$'s. In the remainder of the paper, we will concentrate on probing the decay in Eq. \ref{stopslepdecay}.

\section{Stop to Slepton Decay} \label{stop2slepanalysis}

Our benchmark scenario will be chosen such that the decay in Eq. \ref{stopslepdecay} is competitive. The benchmark points are chosen with the criterion that it is compatible with all existing searches and gives the required value of $(g-2)_{\mu}$, following the right panel of Fig. \ref{Binoslepton2}. The benchmark spectrum is given in Table \ref{benchparameters0}. The case of $\Delta M \, = \, (m_{\st} - m_{\neu{1}}) \, = \, 80$ GeV, which will turn out to be our best-case scanerio, is shown.
\begin{table}[!htp] 
\caption{Relevant SUSY masses (in GeV) at  benchmark point. $\tan{\beta} = 36$, and $\Delta a_\mu = 25.9 \times 10^{-10}$. We keep $\neu{2} \sim \chpm{1} = 748$ GeV, $\tilde t_2 = 730$ GeV, $\tilde{b}_1 = 760$ GeV, $\mu = 3.3$ TeV. All other colored and non-colored states are heavy.}
\label{benchparameters0}
\begin{center}
\begin{tabular}{c c c} \hline \hline
 Particle  & Mass (GeV)      & ${\rm Br}$  \\ \hline  \\[-.1in]
  $\tilde t_1$  & $348$     &  $38\%$ ($b \ell \tilde{\nu}_\ell$), $62\%$ ($b \tilde{\ell} \nu_\ell$) \\
$\tilde{\ell}$ & $302$       & $100\%$ ($\ell \neu{1}$)  \\
$\tilde{\nu}$ & $290$       & $100\%$ ($\nu \neu{1}$)  \\
$\tilde{\ell}_R$ & $594$       & $100\%$ ($\ell \neu{1}$)  \\
  $\neu{1}$  & $270$  & \\    \hline \hline

\end{tabular}
\end{center}
\end{table}

The challenges of probing the decay mode in Eq. \ref{stopslepdecay} with the spectrum of Table \ref{benchparameters0} are symptomatic of compressed stop searches in general: the lack of $\met$ (near the compressed limit, the $\neu{1}$ provides little transverse missing energy) and the softness of the final state $b$-jets, which makes $b$-tagging difficult. Indeed, the current constraints on stops are feeble in the compressed regions, where $ 175$ GeV $ > (m_{\st} - m_{\neu{1}}) \, \gsim $ 85 GeV with dominant decay mode $\st \rightarrow b W \neu{1}$, and  $ 85$ GeV $ \gsim (m_{\st} - m_{\neu{1}}) \, \gsim $ 0 GeV, with dominant decay mode $\st \rightarrow b W^{(*)} \neu{1}$. The decay chain in Eq. \ref{stopslepdecay} has the advantage of full branching to leptons.

\subsection{Results and Analysis}

Although the current run of the LHC is at 13 TeV, we carry out our analysis at 14 TeV, since our sensitivity is only at very high luminosity. Inclusive $\st \st^{\ast} \, + \,$ jets samples are generated with $\st$ masses in the range of $200-400$ GeV, choosing several different values of $\neu{1}$ and hence $\Delta M$. The left handed sleptons are kept at a mass diference $\sim 30$ GeV above the $\neu{1}$. 
The $\neu{1}$ is mostly Bino. 

The spectrum is generated with \suspect \, \cite{Djouadi:2002ze} . Signal and background samples are generated with \madgraph \, \cite{Alwall:2011uj} followed by the parton showering and hadronization with \pythia \, \cite{Sjostrand:2006za} and the detector simulation using \delphes \, \cite{deFavereau:2013fsa}. The study is performed in the $2l + 2b + \met$ final state. The simulated backgrounds are $\ttbar \, + \, (0-2) j$, $V \, + \, (1-3) j$, and $t \, + (1-3) j$, where $V$ stands for $W$ and $Z$. We have used backgrounds from the Snowmass Energy Frontier Simulations \cite{Anderson:2013kxz}.

The following selections are applied:

$(1)$ Preselection: Two isolated leptons with $p_T \geq 10$ GeV and two $b$-jets with $p_T \geq 30$ GeV in $|\eta| < 2.5$ are required. 
The $b$-tagging efficiency is taken to be the default \delphes \, value.

\begin{figure*}[!t]
\centering
\mbox{\includegraphics[width=3.5in]{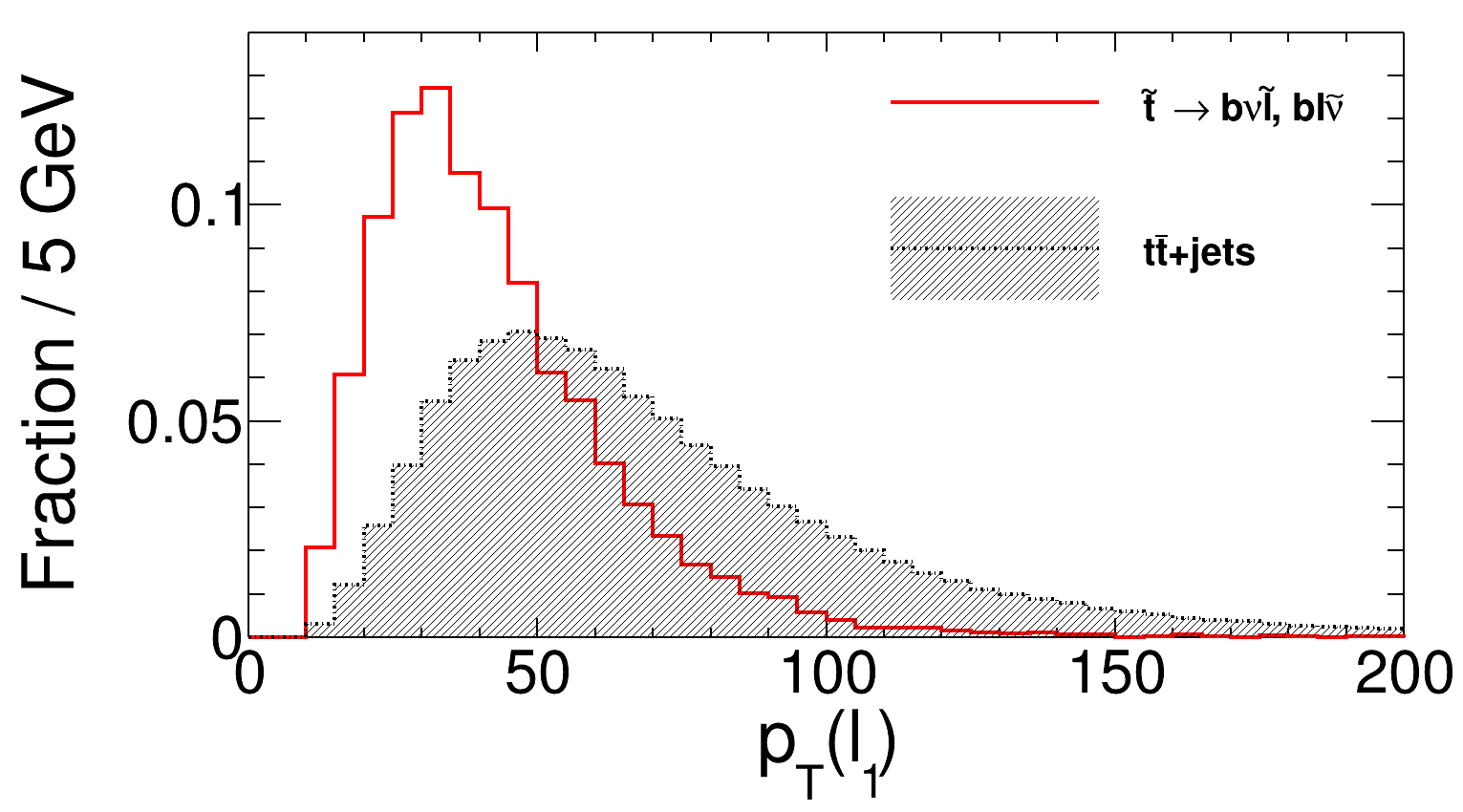}\quad\includegraphics[width=3.5in]{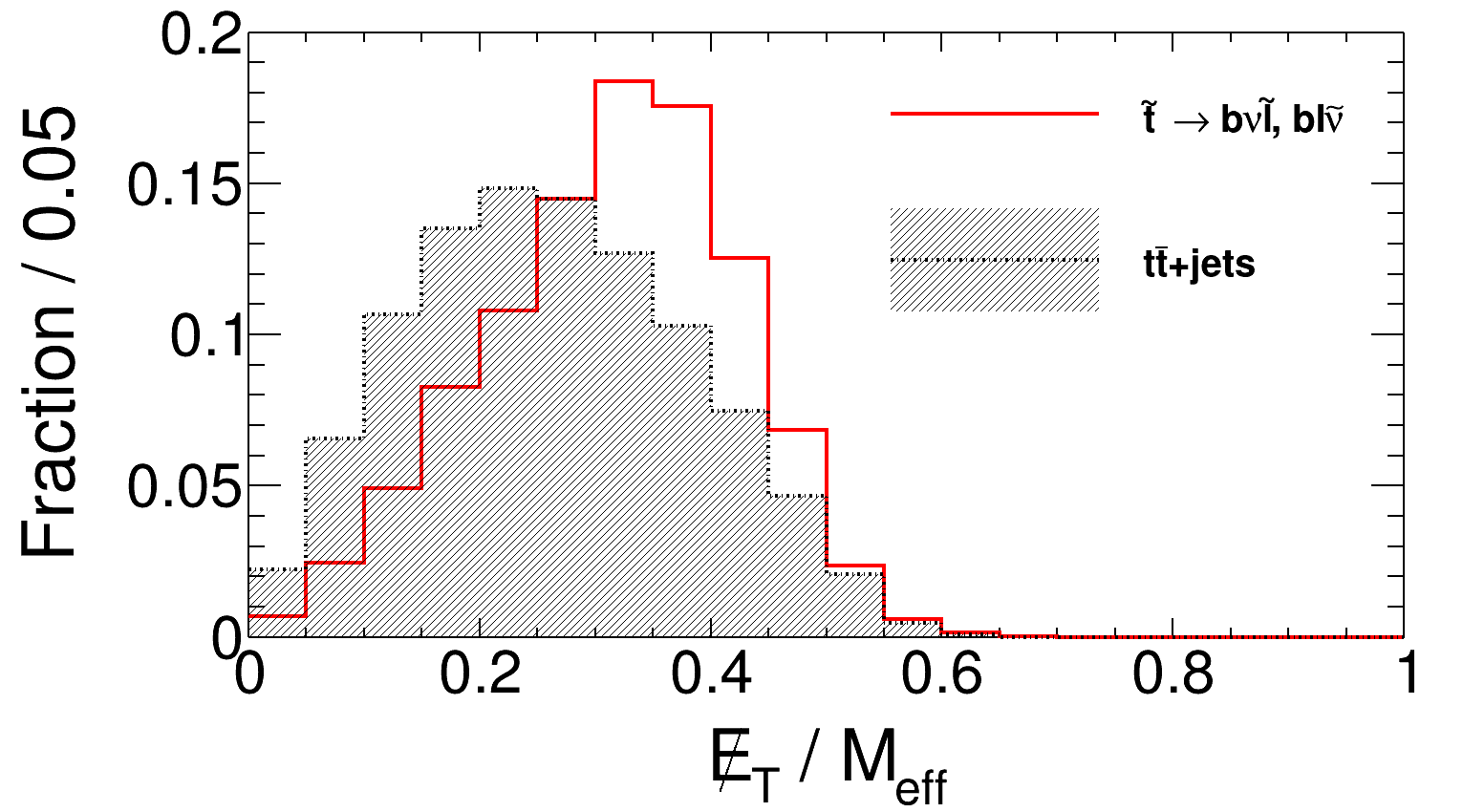}}
\caption{The left panel shows the $p_T$ distribution (normalized to unity) of the leading and next-to-leading leptons for signal (red unshaded histogram) and background (grey shaded histogram) after preselection cuts. The right panel shows  the distribution (normalized to unity) of $\met / M_{eff}$  for signal (red unshaded histogram) and background (grey shaded histogram) after lepton selections. All masses are in GeV.}
\label{stopslepL}
\end{figure*}

$(2)$ Lepton $p_T$: The leading lepton is required to have $p_T(\ell_1) \, < \, 50$ GeV, while the next-to-leading lepton is required to have $p_T(\ell_2) \, < \, 25$ GeV. These selections are based on Fig.~\ref{stopslepL}, left panel.

$(3)$ Events are required to satisfy $\met / M_{eff} \, > \, 0.25$. Here, $M_{eff}$ denotes the scalar-summed transverse momenta of jets and $\met$. This selection is based on Fig.~\ref{stopslepL}, right panel.

\begin{figure*}[!t]
\centering
\mbox{\includegraphics[width=3.5in]{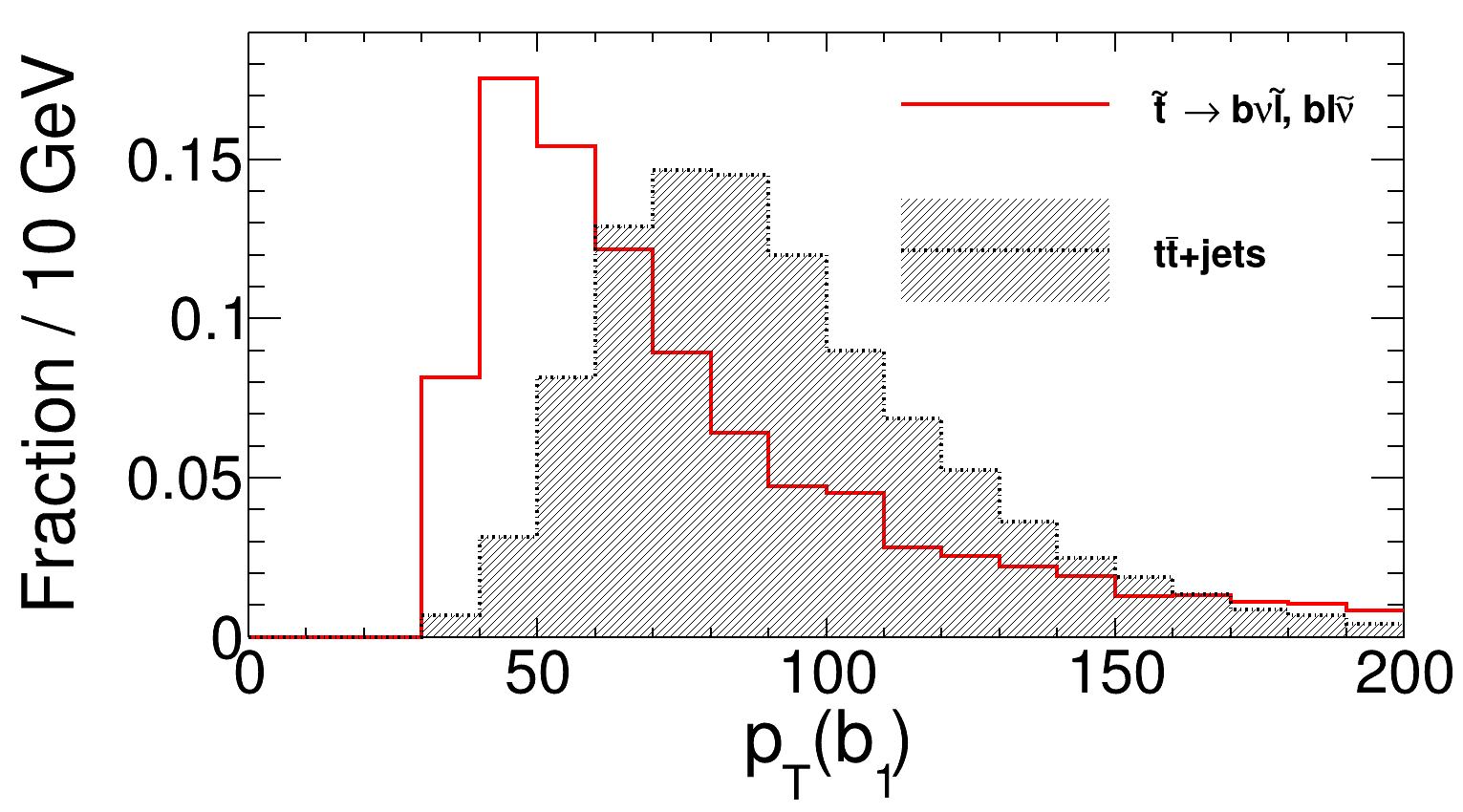}\quad\includegraphics[width=3.5in]{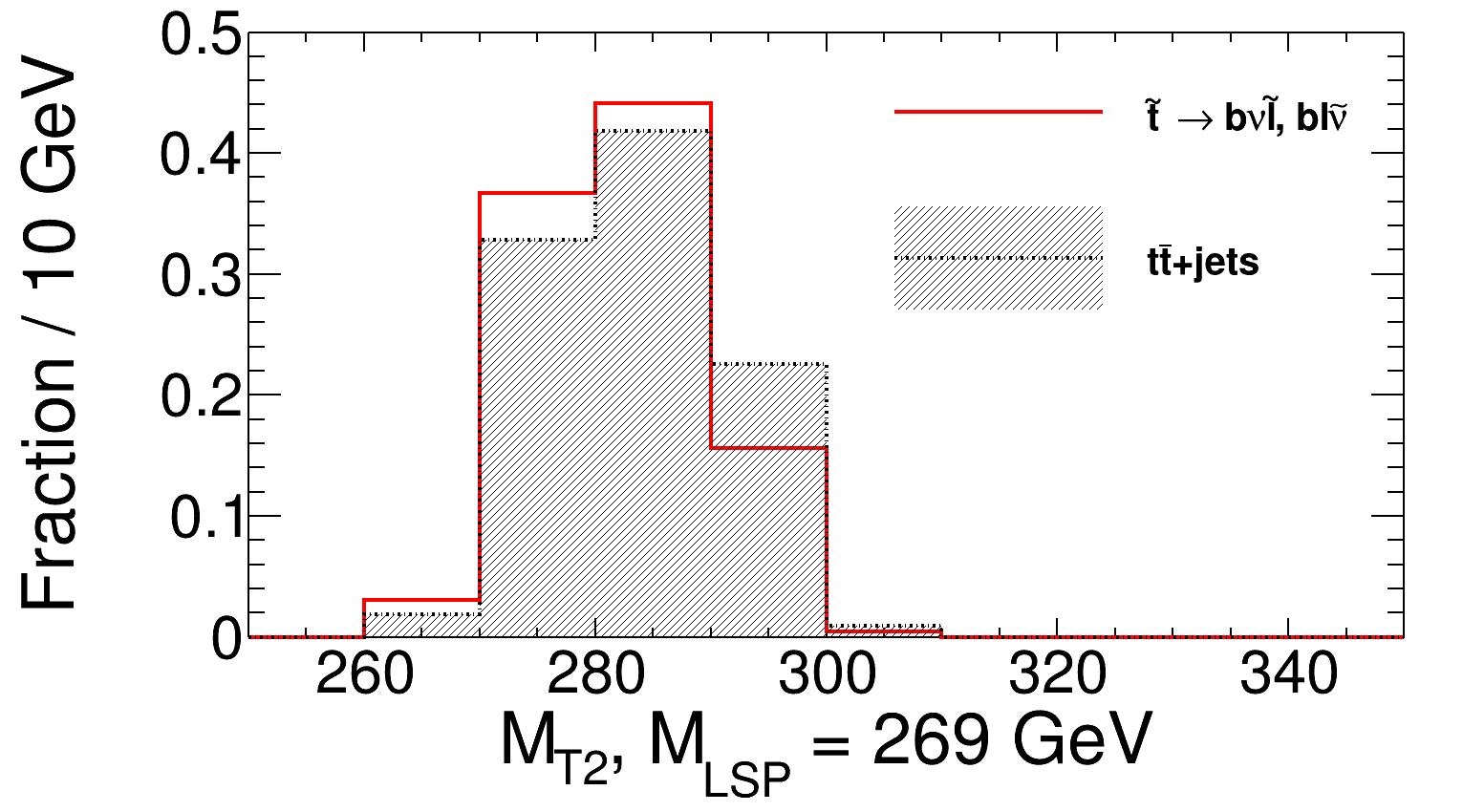}}
\caption{The left panel shows the $p_T$ distribution (normalized to unity) of the leading $b$-jet for signal (red unshaded histogram) and background (grey shaded histogram) after the $\met / M_{eff}$  selection. The right panel shows the distribution (normalized to unity) $m_{T2}$ (right panel) for signal (red unshaded histogram) and background (grey shaded histogram), after all other selections. All masses are in GeV.}
\label{stopslepB}
\end{figure*}

$(4)$ $b$-jet $p_T$: The leading $b$-jet is required to have $p_T(b_1) \, < \, 60$ GeV, while the next-to-leading $b$-jet is required to have $p_T(b_2) \, < \, 40$ GeV. These selections are based on Fig.~\ref{stopslepB}, left panel.

$(5)$ Events are required to satisfy $m_{T2} \, < \, 290$ GeV, for an LSP selection of 270 GeV. This selection is based on the right panel of Fig.~\ref{stopslepB}. We have used the following definition of $m_{T2}$
\bea
&& m_{T2}(\mathbf{p}_{T \ell_1}, \mathbf{p}_{T \ell_2}, \overrightarrow{\met}) = \nonumber \\  
&& min (max (m_T(\mathbf{p}_{T \ell_1}, \mathbf{p}^{miss}_{T,1}), m_T(\mathbf{p}_{T \ell_2}, \mathbf{p}^{miss}_{T,2}))) \,\,,
\eea
with the minimization carried out over all decompositions of $\overrightarrow{\met}$ subject to $\mathbf{p}^{miss}_{T,1} + \mathbf{p}^{miss}_{T,2} = \overrightarrow{\met}$. Here, $m_T$ denotes the transverse mass, while $\mathbf{p}_{T \ell_1}$ and $\mathbf{p}_{T \ell_2}$ denote the transverse momentum vectors of the selected leptons. $\overrightarrow{\met}$ is the missing momentum vector. In the calculation, the mass of missing momentum is chosen to be the mass of $\neu{1}$ for each signal point.

In Fig.~\ref{stopslepL}, left panel, we show the $p_T$ distribution (normalized to unity) of the leading and next-to-leading leptons for signal (red unshaded histogram) and background (grey shaded histogram) after preselection cuts. The right panel shows  the distribution (normalized to unity) of $\met / M_{eff}$  for signal (red unshaded histogram) and background (grey shaded histogram) after lepton selections.

In Fig.~\ref{stopslepB}, left panel, we show the $p_T$ distribution (normalized to unity) of the leading $b$-jet for signal (red unshaded histogram) and background (grey shaded histogram) after the $\met / M_{eff}$  selection. For the signal, the leading  lepton and $b$-jet transverse momenta are approximately determined by the mass difference between $(\tilde{\ell}, \tilde{\nu})$ and $\neu{1}$, and $\st$ and $(\tilde{\ell}, \tilde{\nu})$, respectively, and are softer than the background values. The right panel shows the distribution (normalized to unity) $m_{T2}$ (right panel) for signal (red unshaded histogram) and background (grey shaded histogram), after all other selections.


The cut flow table with corresponding cross-sections at each stage are shown in Table \ref{tablestopbenchmark} for the benchmark point. We only display $\ttbar + j$ background, although we have taken into account all the backgrounds mentioned previously, which are subdominant. After all the cuts, the cross section of $t + (1-3)$ jets is 0.023 fb; the cross section of $V + (1-3)$ jets is 8.4 fb.

\begin{table}[!htp] 
\caption{Summary of the effective cross-sections (fb) for different benchmark signal points as well as the total background at LHC14. Masses and momenta are in GeV.}
\label{tablestopbenchmark}
\begin{center}
\begin{tabular}{c c c} 
\hline \hline 

      Selection      &Signal    &\,\, $\ttbar + j$    \\
          
\hline 
              \hline  \\
                     
                  Preselection  & 7.4   & 16888   \\
         Lepton $p_T$   & 4.2    & 3251   \\
          $\met / M_{eff} \, > \, 0.25$    & 3.1 & 1697   \\ 
        $b$-jet $p_T$       & 0.9 & 87.5   \\ 
                               $m_{T2} \, < \, 290$    & 0.74 & 66.9   \\     
                         \hline \hline \\

\end{tabular}
\end{center}
\end{table}

The significance $S/\sqrt{S+B}$, where $S$ and $B$ are the signal and background rates, respectively, is plotted in Fig. \ref{sigplot} as a function of $m_{\st}$ and $\Delta M$, for $3000$ fb$^{-1}$ of integrated luminosity at LHC14. The blue regions show the $3 \sigma$ reach, while the red regions show the $5 \sigma$ reach. The highest reach is obtained for $\Delta \, M \, \sim \, 80$ GeV. For larger $\Delta \, M$, the significance drops due to reduced branching to leptons, as the dominant decay mode reverts back to the usual $\st \, \rightarrow \, b W \neu{1}$.  For smaller $\Delta \, M$, the $b$-jets coming from stop decay become too soft for the signal events to pass the preselection cuts.

\subsection{Comments on Systematics}

\begin{figure}[!htp]
\centering
\includegraphics[width=3.5in]{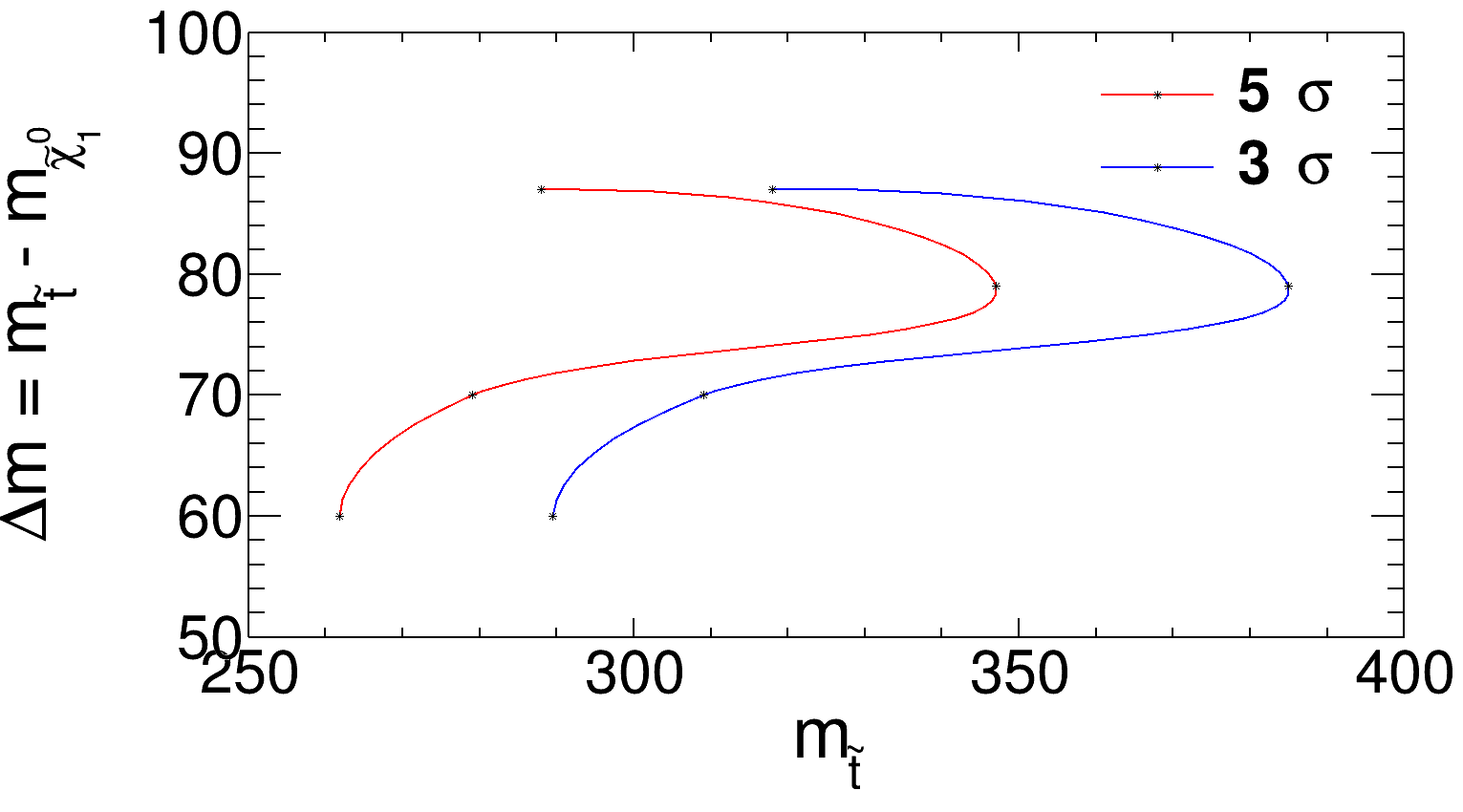}
\caption{The significance $S/\sqrt{S+B}$, where $S$ and $B$ are the signal and background rates, respectively, as a function of $m_{\st}$ and $\Delta M$, for $3000$ fb$^{-1}$ of integrated luminosity at LHC14. The blue contour shows the $3 \sigma$ reach, while the red contour shows the $5 \sigma$ reach. All masses are in GeV.}
\label{sigplot}
\end{figure}

The mass reach for our study shown in Fig. \ref{sigplot} does not take into account the effect of systematics. We now make some comments in this regard.

\begin{itemize}

\item Clearly, one of the most crucial factors determining signal discrimination is the identification of $b$-jets, as part of the preselection cuts. This requires efficient and robust $b$-tagging at low $p_T$. In our study, we have taken the conservative $b$-jet $p_T$ threshold of 30 GeV. The challenge of $b$-tagging at low $p_T$ is expected to be more difficult in the pileup conditions of 14 TeV, although preliminary detector upgrade studies have shown the ability to go down to $p_T = 30$ GeV \cite{cmsbjet2}. In the 8 TeV run, both CMS and ATLAS have shown the ability to identify $b$-jets down to $p_T \, = \, 20$ GeV \cite{cmsbjet1}; this would improve the mass reach for compressed stop searches substantially.

\item Systematic uncertainties are expected to lead to substantial degradation of the mass reach. The background estimation and signal extraction would largely depend on the upgraded detector configuration and trigger conditions at 3000 fb$^{-1}$, and the final systematics in the high pile-up environment would depend on the ability to reject pile-up.

\item The distributions in Fig.~\ref{stopslepL} and Fig.~\ref{stopslepB} show that significant improvement can be achieved by pursuing a shape based analysis. Although we do not perform such an analysis here, we give some estimates. Our $S/B$ for the benchmark point in Table ~ \ref{tablestopbenchmark} is similar to the $S/B$ in a compressed stop study performed by some of the current authors \cite{Dutta:2013gga}, Table II. In that study, a shape based analysis of the $\met$ distribution was performed using a binned likelihood following the test statistic based on the profile likelihood ratio, with systematic uncertainties being incorporated via nuisance parameters following the frequentist approach. Based on that study, we can estimate that the combination of a shape based study and incorporation of $3 \%$ systematics would lead to a reduction of $\sim \, \mathcal{O}(100)$ GeV in the mass reach in Fig.~\ref{sigplot}, which was conducted without the shape analysis and without systematics. Thus, we can expect a $5 \sigma$ reach of around $m_{\st} \, \sim \, 240$ GeV, for $\Delta M \sim 80$ GeV. We leave a more detailed study incorporating shape analysis for the future.

\end{itemize}

\section{Conclusions}

The anomalous magnetic moment of the muon is one of the most precisely measured quantities in particle physics. The upcoming New $g-2$ Experiment at Fermilab will measure it with even greater accuracy, further establishing the discrepancy between the SM prediction and experimental value. Simple but powerful order-of-magnitude calculations of the effects of new physics show that supersymmetry is a prime contender in the resolution of this discrepancy; the MSSM indeed provides ample parameter space in this regard.

We have studied the regions of supersymmetric parameter space preferred by the muon $g-2$ anomaly in the context of constraints arising from the first run of the LHC. We have found that the main constraints arise from slepton searches in the dilepton channel, and chargino searches in the trilepton channel. A combination of these searches rules out a substantial part of the parameter space for $\tan {\beta} \, \sim \, \mathcal{O}(10)$. Most of the parameter space at larger $\tan{\beta} \, \sim \, \mathcal{O}(40)$ remains unexplored, and the constraints here will depend on the future reach of slepton and chargino searches in the high-luminosity LHC.

Since a primary motivation for supersymmetric searches at the LHC is the naturalness paradigm, we have evaluated the allowed parameter space for the muon $g-2$ anomaly in the light of stop-assisted chargino and neutralino production. We have found that current constraints coming from stop searches in the $\st \rightarrow t \neu{1}$ and $\st \rightarrow b \chpm{1}$ are already probing the entire parameter space at $\tan{\beta} \, \sim \, \mathcal{O}(10)$, and substantial parts of it at $\tan{\beta} \, \sim \, \mathcal{O}(40)$. 

The future performance of the LHC in these stop searches will be crucial in illuminating the interplay between naturalness and a supersymmetric resolution to the muon $g-2$ anomaly. One possible blind spot are the compressed stop scenarios. We have studied one such scenario, which can be dominant in the presence of light charginos and sleptons in the parameter space compatible with the muon $g-2$ anomaly: $\st \, \rightarrow \, b \tilde{\ell} \nu (\tilde{\nu} \ell) \, \rightarrow \, b \ell \nu \neu{1}$, where $\Delta M \, = \, m_{\st} - m_{\neu{1}} \, \sim \, 80$ GeV. Using a simple cut and count approach, we have found that the $5\sigma$ mass reach in such a scenario is $m_{\st} \, \sim \, 350$ GeV. Systematics will degrade this reach substantially; to obtain sensitivity, we have pointed out that shape analysis of kinematic distributions will be essential to discriminate between signal and background. This will further probe the regions of paameter space where  naturalness and a supersymmetric resolution to the muon $g-2$ anomaly are compatible.

\section{Acknowledgements}

B.P.P. is supported by DOE Award DE-SC0010103. K.S. is supported by NASA Astrophysics Theory Grant NNH12ZDA001N. K.W. is supported in part by the CAS Center for Excellence in Particle Physics (CCEPP).  K.W. is grateful to Cai-Dian Lu and wants to thank Masaaki Kuroda, Jinrui Huang and Lian-Tao Wang for useful discussions.

\end{document}